\documentclass[11pt,twoside]{article}

\usepackage{asp2004}
\usepackage{graphicx}

\begin{document}
\title{Echoes of Giant Pulses from the Crab Pulsar}   
\author{J. H. Crossley, J. A. Eilek and T. H. Hankins}   
\affil{New Mexico Tech, Socorro NM, USA}    

\begin{abstract} 
We have detected occasional, short-lived ``echoes'' of giant pulses
from the Crab pulsar.  These echo events remind us of  previously
reported echoes from this pulsar, but they differ significantly in
detail. Our echo events last at most only a few days;  the echo emission lags
the primary emission by only $40\!-\!100\ \mu$s.  The echoes 
are consistently weaker and broader than the primary emission, and
appear only at the lower of our two simultaneous observing
frequencies.  We suggest that these echoes are created by refraction
in small plasma structures --- plasma clouds or magnetic flux ropes ---
deep within the Crab nebula.  If this is true, our echoes provide a
new probe of small-scale structures within the inner synchrotron
nebula.   
\end{abstract}

Our group has been working to understand the physics of the pulsar
emission region.  To that end, we have pursued high time resolution
observations, because differences in the physics should lead to
different time signatures in the radio emission. 
We have concentrated on the Crab pulsar, because its
occasional very bright ``giant'' pulses are well suited to our data
acquisition systems.  The results we report on here come from a
VLA program to observe single pulses from the star at sub-$\mu$s time
resolution. 

The pulsar sits at the heart of the Crab
Nebula --- which is, of course, powered by spindown of the pulsar. The
complex nebular structure includes the relativistic 
pulsar wind;  the shock ($\sim 0.1$ pc from the pulsar) at which the
cold wind is thermalized;  the highly variable ``wisps'' and
``knots'' associated with the dynamic, post-shock region (Hester
et al.\ 2002;  Bietenholz et al.\ 2004);   
the diffuse synchrotron nebula ($\sim 0.1\!-\!1$ pc from the pulsar),
which is thought to contain shocked wind plasma; and the cool,
filamentary outer regions (emiting optical emission lines and radio
bremsstrahlung), which may be due to  Rayleigh-Taylor
instabilities at the interface between the synchrotron nebula and the
stellar ejecta. 

Because our interest was in the pulsar itself, we  initially saw
the nebula mostly as a nuisance.  Its high sky brightness makes
observations of fainter pulsar signals challenging; its internal
``weather'' has long been known to cause fluctuations in dispersion,
Faraday rotation and scattering broadening of low-frequency pulses.
We were surprised, then, to notice that our high-frequency, high-time
resolution observations had detected what we think is a new type of
nebular weather:  small (AU-size), dense fluctuations within the
nebular synchrotron plasma.  In this paper we summarize our
observations and our suggested interpretation;  we give more details 
in Crossley, Eilek \& Hankins (2007).

\section{Observations:  echoes of the giant pulses}

We used the VLA to record single pulses from the Crab 
pulsar on 20 observing days
between 1993 and 1998, with a 50-MHz bandwidth  centered
at 1.4 or 4.9 GHz.  Some days we split the VLA into 2 subarrays to
record both frequencies 
simultaneously.  A typical giant pulse contains one to several
short-lived microbursts, which last 
$\sim 10\  \mu$s at 1.4 GHz, and are shorter at 4.9 GHz (e.g., Hankins
2000).   These microbursts generally seem to be random, in rotation
phase and amplitude, suggestive of unrelated flare events in the radio
emission region. However, a different pattern emerged on two observing
days.  A bright ``primary'' burst at 1.4 GHz is followed {\it in
  nearly every recorded pulse}\/ by a fainter, broader ``echo'' burst;
the same pulse at 4.9 GHz shows no echo.  Figure 1
shows three example pulses. The echo properties,  relative to the
primary, remained constant for the full duration of each day's
observing run (32 clear echoes within 3 hours on one echo-event day; 9
clear echoes within 15 minutes on the other day). 
The unusual consistency of the
echo properties during each observing day  leads us to suggest these
echoes are not intrinsic to the star, but rather are due to some chance
structure in the nebular plasma which happened to lie along the line of sight
on those  days.

\begin{figure}[htb]
\centerline{\rotatebox{-90}{
\includegraphics[width=0.7\columnwidth]{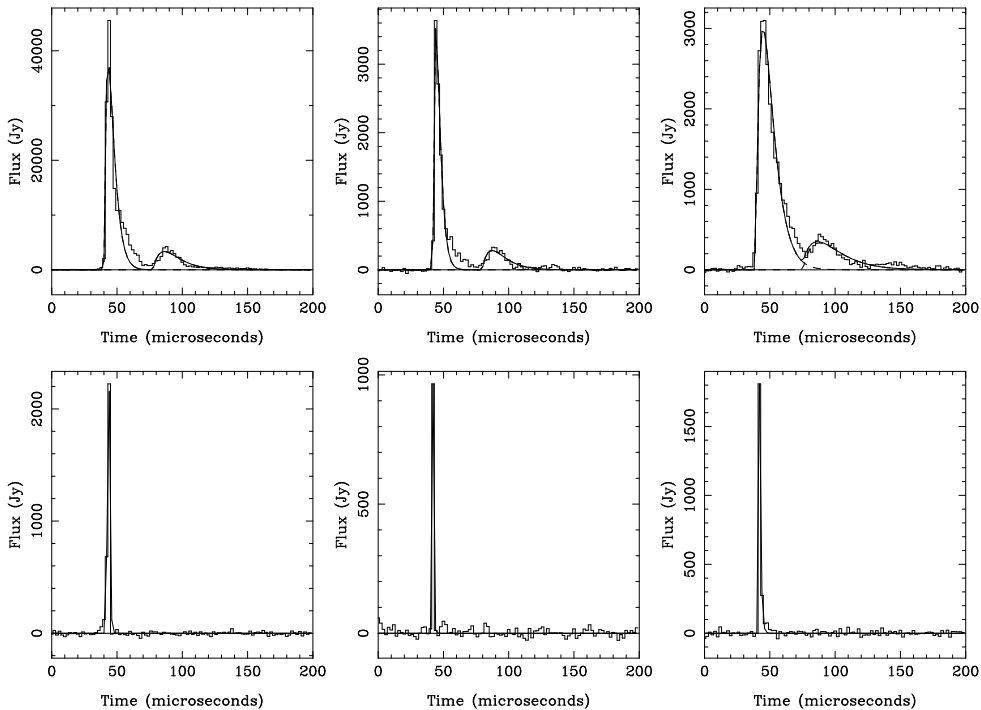}}}
\caption{Three giant pulses from the Crab pulsar, recorded
 simultaneously at 1.4 and 4.9 GHz within two hours at the VLA, and
 displayed with 2-$\mu$s time resolution.  The top row shows  the
 pulses at 1.4  GHz;  echo emission following  the primary pulse 
 by $\sim 50\ \mu$s is  apparent.  The bottom row shows the same three
 pulses, but now at 4.9  GHz, where no echoes are seen.  The smooth, light
 lines on the 1.4-GHz pulses are the component fits we
 used to measure arrival times, widths and energies of pulse
 components.} 
\end{figure}

In order to quantify our results, we fit individual microbursts 
 with the  function $I(t) = A
(t-t_o) e^{-(t-t_o)/\tau}$. 
This enabled us to determine the component arrival time ($t_o$), width
($\propto \tau$), peak flux ($\propto A \tau$), and energy content ($E
= A \tau^2$).  The echo properties we derive are remarkably uniform
within each echo-event day.  The echoes are broader: $\tau_{\rm echo}/
\tau_{\rm primary} \sim 
2.5$ on one echo-event day, $\sim 4.6$ on the other day.  The echoes are
fainter:  the peak echo flux $\sim 10\%$ of the peak primary flux, on
both echo-event days.  Their energy ratio is
 $E_{\rm echo}/E_{\rm primary} \sim
0.25$ on one day, and $\sim 0.35$ on the other day.   Their time lag,
relative to the primary, is $\sim 45\ \mu$s on one day, $\sim 90\ \mu$s
on the other day.   On both echo-event days, we see the echoes clearly
at 1.4 GHz, but not at 4.9 GHz.   
We also observed the pulsar two days before and two
days after the first echo-event day, without seeing any echoes.  We
can thus bracket the duration of this echo event:  it lasted between
three hours and four days.

\section{Interpretation:  small structures in the Crab Nebula}

Our  echoes are reminiscent of similar events reported by other authors.
Lyne, Pritchard \& Graham-Smith (2001) reported 10 such events,
detected in time-averaged mean pulse profiles at 610 MHz.  One
particularly dramatic event was analyzed in detail by Lyne et al.\
(2001), and also by Backer, Wong \& Valanju (2000).  If this 
particular echo event is characteristic of all reported by Lyne et al., 
these previous echo events differ quantitatively from ours.  They 
last much longer (several tens of days), during which the echo lag
changes smoothly between several milliseconds and the time resolution of the
observations ($\sim 1$ ms).  The echoes often appear to approach the primary,
then recede from it.  These previous  echoes are achromatic;  they have
the same characteristics at 327 and 610 MHz (reported by Backer et al.\
2000).  At least one echo event coincided with a significant
jump in dispersion measure, $\Delta {(\rm DM)} \sim 0.1$cm$^{-3}$pc.
Backer et al.\ (2000) and Lyne et al.\ (2001) modeled this event as 
refraction or reflection from a dense plasma structure
related to the cool filaments in the outer regions of the Crab nebula.

By contrast, our echo events are shorter-lived, have a much shorter
time lag, and are frequency dependent.  Can they  be explained by
something similar?  We first note that  
the $\sim 50\!-\!100$-$\mu$s echo lag must be due to a longer
propagation path. If the time lag were dispersive, $\propto 1 /
\nu^2$, we would expect to see similar strength echoes with 
4 to 8-$\mu$s lags at 4.9 GHz. No such echoes were found, either in direct
inspection of individual pulses, or in autocorrelation analysis.   We
therefore conclude that the echo time lag is geometrical, due to the echo 
path length being   $\sim 15\!-\!30$ km longer than that of the primary.

Because the time delay is geometrical, and the echo is only seen at
the lower frequency, we hypothesize that the 1.4-GHz echo is due to 
reflection from some plasma structure within the nebula. We envision
this structure as a dense plasma cloud in a turbulent region, or
possibly dense plasma within a magnetic flux rope.   We note that  
``reflection'' in this context should be interpreted as total internal
reflection, caused by refraction when the signal propagates into a
higher density plasma.  The lack of any echo at 4.9 GHz must mean the
density in the plasma structure is not high enough to reflect that
frequency.   If the reflecting structure is
at a distance $R$ from the pulsar, and the beam deflects by angle
$\theta$, a 50-$\mu$s lag requires  $\theta^2 R \sim 30$ km. We do not
know $R$, but if our feature is associated with the nebular
synchrotron plasma, we expect $0.1 ~{\rm pc} < R < 1 ~{\rm pc}$.
If the duration of the echo event is due to a plasma cloud moving
across the line of sight, the transverse size of the cloud must be
no more than an AU. 

We also recall that full reflection occurs if the incident angle,
$\delta$, 
relative to the surface of the reflecting structure, satisfies $\delta
\leq \nu_p / \nu$ (where $\nu_p$ is the plasma frequency;  this
assumes the cold  plasma dispersion law, and small angle
approximations). If we  combine this condition with the 
time delay, we can bracket the density of the reflecting feature:  $0.003~
{\rm cm}^{-3}\!<\!n\!<\!0.4~{\rm cm}^{-3}$.  (If the plasma is internally
relativistic, with mean particle energy 
$\gamma m c^2$,  $n$ in our result should be replaced by $n / \gamma $;
e.g., Wills \& Cairns 2000).  The other properties of our echoes also
seem to fit this picture.  If the plasma within the
reflecting structure is itself turbulent, multipath propagation will broaden
the echo relative to the primary.  If the surface of the reflecting
structure has some curvature, dispersion of the reflected beam will
cause the echo to be weaker than the primary.

Can our speculation be tested?  The density range we require for such
plasma clouds  is well below that 
estimated for the cool, outer filaments.  Where might AU-sized features
with this density range exist within the Crab Nebula?
We think the most likely place to find such features is the diffuse
synchrotron nebula, which is almost certainly  turbulent and
inhomogeneous (probably filamented into disordered 
magnetic flux ropes, as Hester et al.\ 1995 point out).  We note that,
in addition to the ordered, elliptical ripples (``wisps'') seen close to the
pulsar,  a fair amount of faint, disordered structure is  apparent
in the images (e.g., Bietenholz et al.\ 2004).  We suspect this is the
``tip of the iceberg'' of the turbulence;  but such  structures are
hard to quantify. The most sensitive observations to date are those 
from HST (Hester et al.\  1995).  If we apply standard (equipartition)
synchrotron analysis to the smallest, highest-emissivity knot they
report,  we find the plasma density in that knot may  just  be
consistent with our required density range.   We also note that our
estimates have  a fair amount of ``wiggle room''.   We fully expect
smaller, denser features to exist below HST's sensitivity and
resolution;  and we recall that equipartition is only an approximation
to the true state of the plasma.  We therefore  conclude that  the
sources of our echoes are likely to be dense, transient  clouds or
flux ropes in the turbulent synchrotron plasma within the Crab
Nebula. 

\acknowledgements 

This work was partially supported by NSF grant AST-0139641.  The VLA
is run by NRAO, a facility of the NSF, operated under cooperative
agreement by Associated Universities, Inc.


\end{document}